\documentclass[10pt,twocolumn]{article}
\usepackage[top=1.5cm, bottom=1.5cm, left=1.5cm, right=1.5cm]{geometry}

\usepackage{paralist}
\usepackage{amssymb}

\usepackage[hyphens]{url}

\usepackage{comment}
\usepackage{listings}
\usepackage[dvipsnames,table,xcdraw]{xcolor} 
  
\usepackage{booktabs} 

\usepackage{graphicx}
\graphicspath{ {figures/}{biopics/} }  

\usepackage{times}
\usepackage{pifont}

\usepackage{enumitem}
\usepackage{subfig}
\usepackage{multirow}
\usepackage{braket}
\usepackage{mathtools}

\DeclarePairedDelimiter{\floor}{\lfloor}{\rfloor}

\usepackage{titlesec}

\let\subsubsection\paragraph
\newcommand{\simicoldafter}[1]{#1:}
\titleformat{\subsubsection}{\normalfont\scshape}{\textbf{\thesubsubsection.}}{1ex}{\simicoldafter}

\def\signed #1 {{\unskip\nobreak\hfil\penalty50
    \hskip2em\hbox{}\nobreak\hfil\sl---#1\/
    \parfillskip=0pt \finalhyphendemerits=0 \par}}

\newcommand{\PublicRepo}{\url{https://github.com/gsuareztangil/automatic-romancescam-digger}}

\usepackage[bookmarks=true,%
            bookmarksnumbered=true,%
            colorlinks=true,%
            linkcolor=linkcol,%
            citecolor=citecol,%
            urlcolor=urlcol,%
            hypertexnames=true,%
            pdfpagelabels]%
      {hyperref}

\definecolor{linkcol}{rgb}{0,0,0.5}
\definecolor{citecol}{rgb}{0,0.5,0.3}
\definecolor{urlcol}{rgb}{0.3,0,0}

\newcommand{\guillermo}[1]{\noindent{\color{Cyan}{\bf \fbox{GST}} {\it#1}}}
\newcommand{\matthew}[1]{\noindent{\color{DarkOrchid}{\bf \fbox{ME}} {\it#1}}}
\newcommand{\claudia}[1]{\noindent{\color{Brown}{\bf \fbox{CP}} {\it#1}}}
\newcommand{\awais}[1]{\noindent{\color{blue}{\bf \fbox{AR}} {\it#1}}}
\newcommand{\gianluca}[1]{\noindent{\color{Magenta}{\bf \fbox{GS}} {\it#1}}}

\renewcommand{\guillermo}[1]{}
\renewcommand{\matthew}[1]{}
\renewcommand{\claudia}[1]{}
\renewcommand{\awais}[1]{}
\renewcommand{\gianluca}[1]{}

\begin{document}

%

\title{Automatically Dismantling Online Dating Fraud\thanks{A shorter version of this paper appears in IEEE Transactions on Information Forensics and Security. This is the full version.}}
%
%
%

\author{Guillermo~Suarez-Tangil$^{\star}$, 
        Matthew~Edwards$^{\ddagger}$, 
        Claudia~Peersman$^{\ddagger}$, \\
        Gianluca~Stringhini$^\dagger$, 
        Awais~Rashid$^{\ddagger}$, 
        and~Monica~Whitty${^+}$\\[0.5ex]
{$^{\star}$King's College London, $^\ddagger$}University of Bristol\\
{$^\dagger$Boston University, ${^+}$University of Melbourne}
}

\date{}

\maketitle

\begin{abstract}
Online romance scams are a prevalent form of 
mass-marketing fraud in the West, and yet few studies have 
addressed the technical or data-driven responses to this problem. 
In this type of scam, fraudsters craft fake profiles and manually 
interact with their victims. Because of the characteristics of this 
  type of fraud and of how dating sites operate, traditional detection methods 
(e.g., those used in spam filtering) are ineffective. In this paper, 
we present the results of a multi-pronged investigation into the 
archetype of online dating profiles used in this form of fraud, 
including their use of demographics, profile descriptions, and images, 
shedding light on both the strategies deployed by scammers to appeal to 
victims and the traits of victims themselves. Further, in response to 
the severe financial and psychological harm caused by dating fraud, 
we develop a system to detect romance scammers on online dating platforms. 

Our work presents the first 
system for automatically 
detecting this fraud. Our aim is to provide an early detection 
system to stop romance scammers as they create fraudulent profiles or 
before they engage with potential victims. Previous research has indicated 
that the victims of romance scams score highly on scales for idealized 
romantic beliefs. We combine a range of structured, unstructured, 
and deep-learned features that capture these beliefs. No prior work has 
fully analyzed whether these notions of romance introduce traits that 
could be leveraged to build a detection system. Our ensemble machine-learning 
approach is robust to the omission of profile details and performs at 
high accuracy (97\%). 
The system enables 
development of automated tools for dating site providers 
and individual users.


\end{abstract}


\section{Introduction}
\label{sec:intro}

The online romance scam is a prevalent form of mass-marketing fraud in
many Western countries~\cite{businessInsider,ScamWatch,forbes,whitty2016online}. 
Cybercriminals set up a false user profile on dating 
websites or similar online platforms (e.g., social networking sites, 
instant messaging platforms) to contact potential victims, posing as 
attractive and desirable partners~\cite{whitty2013scammers}. Once contact has
been established, scammers apply a range of techniques to exploit their victims.
In many cases, they engage in a long-term fictitious romantic relationship
to gain their victims' trust and to repeatedly defraud them of large sums of money~\cite{huang2015quit}.

Recently, the FBI~\cite{IC32014} reported a total loss of \$85 million through online romance scams in the US. 
On an individual level, IC3 complaint data showed that 
an average of \$14,000 was lost per reported incident of online dating fraud.
Furthermore, many victims find it difficult to seek support due to being left traumatized by the loss 
of the relationship, and suffer from the stigma of being an online dating fraud victim~\cite{whitty2016online}. 

Despite the magnitude of this type of cybercrime, there is an absence 
of academic literature on the practical methods for detecting romance 
scammers. Previous work has mentioned that online dating sites are 
employing both automated and manual mechanisms to detect fake accounts, 
but do not discuss the specifics~\cite{huang2015quit,edwards2018conPro}. 
Some dating sites are known to use static information such as 
blacklists of IP addresses or proxies to identify alleged 
scammers~\cite{RomancescamIP}. However, these countermeasures 
can easily be evaded through e.g., low-cost proxy services using
compromised hosts in residential address spaces. \matthew{do we have a citation to add here?}
Dealing with online dating fraud is challenging, mainly because such 
scams are not usually run in large-scale campaigns, nor are they 
generated automatically. As a result, they cannot be identified by 
the similarity-detection methods used for spam filtering. Dating websites are designed for connecting strangers and meeting new people, 
which renders the concept of \textit{unsolicited} messages---a key 
element of most state-of-the-art anti-spam systems---strategically 
useless~\cite{huang2015quit}. Finally, romance scammers will send 
a series of ordinary, personalized communications to gain their victims' trust. 
These communications highly resemble messages between
genuine dating site users. In many cases, the actual scam is performed after 
a few weeks or months and after communication has moved to other, unmonitored
media~\cite{whitty2013scammers}. Therefore, it is essential to identify romance 
scammers before they strike. 


Given that the online dating profile is the launching point for the scam, 
it is important to learn a) how scammers craft profiles to draw in potential
victims and b) if there are any distinguishing features of these profiles which
can be identified for automatic detection. This is a distinct problem from the 
detection of Sybil attacks or cloned profiles~\cite{Ramalingam2017fake}, 
existing methods for which typically rely upon graph-based defences or 
markers of automated behaviour, neither of which are applicable here.
Previous research has indicated 
that the victims of romance scams score highly on scales for idealized romantic
beliefs~\cite{buchanan2014online}. Thus, a scammer profile might be expected to exploit these
notions of romance when designing their dating profile. To the best of our knowledge, 
no prior work has analyzed how these notions of romance appear as traits in dating
profiles, or whether these traits can be leveraged to build a scammer detection system. 

In this paper, we present a machine-learning solution that addresses the detection of online dating fraud in a fully automated fashion, which is widely applicable 
across the dating site market---including by the users themselves. 
More specifically, we 
combine advanced text categorization and image analysis techniques to extract useful information 
from a large dataset of online dating user profiles and to automatically identify scammer profiles. 
The key contributions of our work are as follows:

\begin{itemize}
	\item We leverage a large public database of romance scammer profiles, in combination
	with a large random sample of public profiles from a matched online dating site to understand the 
	characteristic distinctions between scammer profiles and those of regular users.
	\item We design three independent classification modules which analyze different aspects
	of public profile characteristics.
	\item We synthesize the individual classifiers into a highly accurate ensemble classification 
	system. Even when parts of the profile information are omitted, our system can reliably  
	distinguish between scammer and real user profiles (\textsc{F1}= 94.5\%, \textsc{Acc}= 97\%), resulting in a solid solution which should generalise well to other dating sites. 
\end{itemize}

To enable replication and foster research we make our tool publicly available at \PublicRepo. 
This paper proceeds as follows. 
In \S\ref{sec:preliminaries} we describe our dataset and the observed characteristics of real and scammer dating profiles. 
In \S\ref{sec:core} we discuss the architecture of our ensemble classification 
system. In \S\ref{sec:evaluation} we detail the division of the data for
training, test and validation purposes, and present our results, before discussing 
related work in \S\ref{sec:background} and concluding with final remarks
in \S\ref{sec:conclusions}.

\section{Characterizing Dating Profiles}
\label{sec:preliminaries}



Though variations exist
within the market, typical dating profiles consist of at least one image
of the user, some basic information about their key attributes, and a 
self-description used as a `sales pitch'. Our approach to scam
detection focuses on these common profile components, which are present across
the market.
In what follows, we compare the characteristics of real dating profiles
with those which are designed by scammers, and detail what we can learn about
scammer targets and strategies from the differences between them.

\subsection{Data}


The data we use comes from a dating site \url{datingnmore.com}, and the connected 
public scamlist at \url{scamdigger.com}. This 
dating site distinguishes itself from the market on the basis of its lack of romance scammers. It  
screens its registrants and members to identify scams, which are then listed
openly to warn the general public and anyone whose likeness may be
being appropriated by the scammers. 
Our dataset combines an exhaustive scrape of the
scammer profiles 
and 
a large random sample of 
one-third of the ordinary dating profiles
as of March, 2017. 
In total, our dataset is composed of 14,720 ordinary profiles, and 5,402 scammer
profiles\footnote{There were roughly 3,500 scammer profiles in the original
data, but these included `or' values where specific attributes which annotators
had seen the profile present differently were given multiple values.  We
exploded these `or' attributes into different profiles to analyze the
profile-variants, but in all analysis that follows were careful to avoid
assigning profile-variants from one original profile to different sets or
folds.}. The sampling of the dating site was spread over a member index sorted
by registration date, to ensure comparison with the scamlist compiled over the
site's operation. 

All data used in this paper is publicly available, with no requirement to register, 
log in or deceptively interact with users of the dating site to collect it. Nevertheless, in
the interests of privacy, no personally identifying information is revealed in
this paper, including that of reported scammers. To enable replication of our
results, we make available two scripts which implement the data-harvesting
process that created our dataset. This enables replication
while allowing dating site users to ``withdraw'' from future study by removing
their profile from public view. The research was approved by the relevant Institutional Review Board (IRB).

The attributes available for scammer profiles and genuine profiles are slightly
different. The scamlist profiles include the IP address, email address and phone
number used by the scammer when registering, along with bookkeeping information
on the justification used for the decision that a profile is a scam. 
These variables are not present in the public member information. Contrarily, there
are attributes visible on public member pages---related to the dating interests of members---which were not duplicated to the scamlist for scammer profiles.
For the purposes of informing discriminative and widely-applicable classifiers, we focus on those attributes which are available for both types of profile, which we divide into the following three groups:

\begin{itemize}
	\item \textbf{Demographics}: Simple categorical information relating to
the user, such as age, gender, ethnicity, etc. 
	\item \textbf{Images}:	One or more images of the user. The dating site mandates that only images showing your own face may be used as an avatar, and 
users are usually motivated to include pictures that illustrate their hobbies. 
	\item \textbf{Description}: A short textual self-description from the user, in which they 
	advertise their key traits and interests.
\end{itemize}

Different techniques are required to extract meaningful information from these profile attributes.
In the following, we cover the preprocessing required for each group, and the notable
features of scammer and real dating profiles.

\subsection{Profile Demographics}
\label{sec:demographics}

\gianluca{I would heavily cut this section. The details about each demographics element are interesting, but we might want to remove some of them}

Dating site demographics act as a filter for users. At the crudest level, most
users will be searching for a particular gender of partner. Typically, age and
other information about a person will also play a role in
their match candidacy. In response to such filtering, users may withhold or
lie about certain demographic characteristics to make themselves seem more
desirable to potential partners. For most real dating site users, any
such deceptions or omissions must be low-level, as they intend for a personal
relationship to result~\cite{hancock2007truth}.
Romance scammers, however, have no expectations of a real relationship, and 
are highly motivated to engage in this form of deception. The
information they present in profiles should thus in no way be taken as an accurate measure of
their true demographics. However, the attributes selected in their profiles
can reveal much about their overall strategies for attracting potential victims 
for romance fraud, and even, implicitly, who their targets may be.

\subsubsection*{Age, Gender, Ethnicity and Marital Status}

The gender distribution of both real and scam profiles was identical: about 60\%
of profiles are male. This highlights that romance scamming is not a
gender-specific problem, in line with the understanding of previous
studies~\cite{whitty2013scammers}.
The average age of real and scam profiles was around 40 in both cases. However,
the distribution of ages differs significantly. Within real profiles, the
average age of male and female profiles is the same, but within scam profiles
the average age of females is roughly 30, and the age of male profiles is
roughly 50. This bimodal distribution around the mean of real profile ages
points at scammer understanding of gendered dating preferences---men here prefer
younger, physically attractive partners, while women prefer partners with
higher socio-economic status, who may be older~\cite{kenrick1990evolution}.


\begin{figure}[t!]
\centering
\subfloat[Profile ethnicities\label{fig:ethnicity}]{\includegraphics[width=.5\columnwidth, trim=0 0.5cm 2.32cm 3cm,  clip]{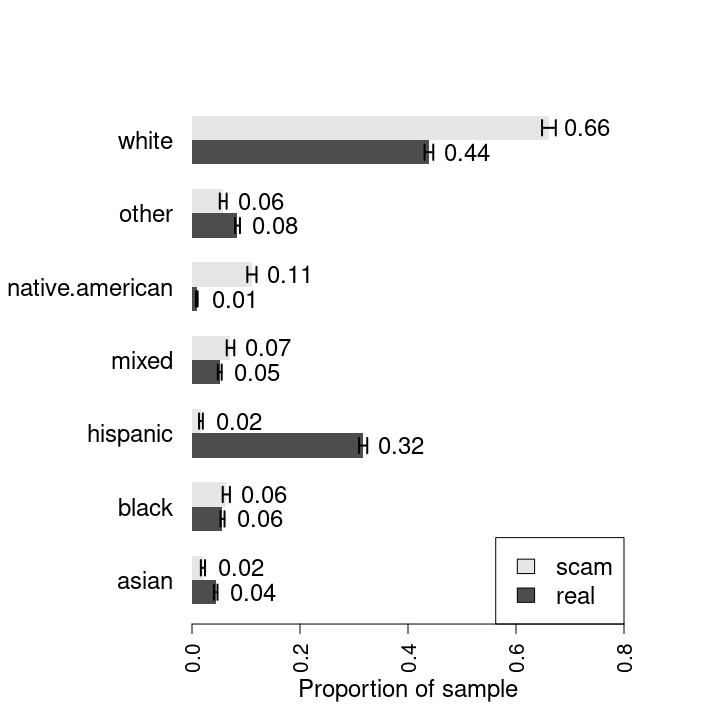}}
\subfloat[Profile marital statuses\label{fig:statuses}]{\includegraphics[width=.5\columnwidth, trim=0 0.5cm 2.32cm 3cm,  clip]{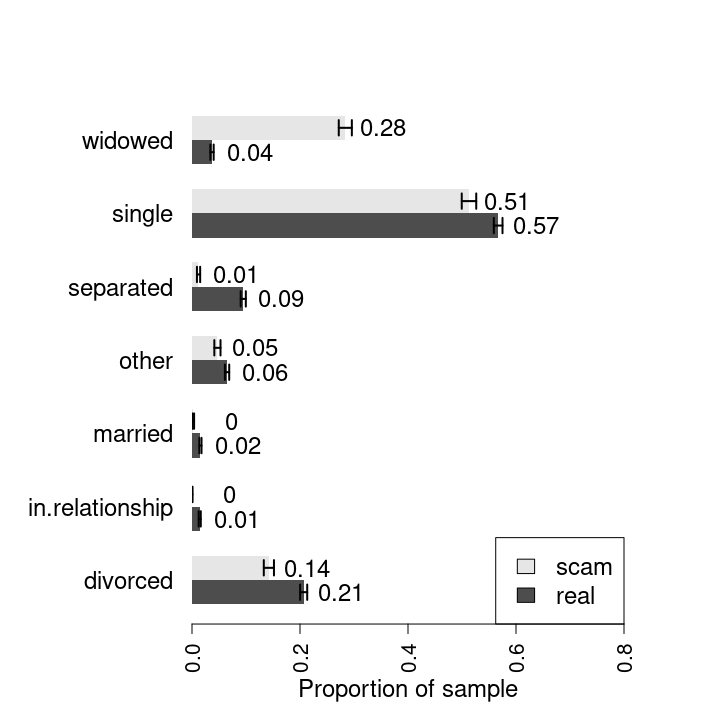}}
\caption{Ethnicity and marital status, with 95\% CIs}
\end{figure}




As reported in Fig.~\ref{fig:ethnicity}, the ethnicities claimed by scammers are 
intriguing. The high proportion which claim to be white is unsurprising, 
as this is the ethnicity of most of their intended victims. However, 
the dating site has a large Hispanic population, 
but the scammers rarely pretend to be Hispanic. Instead, the second most popular
ethnicity amongst scam profiles is Native American, a very small population
amongst the real data. This would seem to reflect some criteria of desirability
which perhaps is related to the fact that dating scams are often targeted at a US-based population.


As Fig.~\ref{fig:statuses} displays, while both real and scam users were
mostly single, scammers prefer to
present themselves as widowed rather than any of the other categories. This is
unsurprising, as female scam victims often talk about such a trait being a
successful strategy to gain their sympathy and trust~\cite{whitty2013scammers}.
Less desirable statuses such as divorced or separated were underrepresented in scam profiles, and 
scammers were far less likely than real users to be married or in a
relationship.

\subsubsection*{Occupation}

There were a wide variety of occupations, 
several being misspellings or rephrases of others.
Responses were grouped into 45 occupation areas. 
Tables~\ref{tab:occupation_male} and \ref{tab:occupation_female} reflect the
major occupation areas for male and female profiles respectively. In both cases,
approx. 15\% of real and 4\% of scam responses were not well-captured by
occupation groupings, this category of `other'
reflecting a long tail of unique occupation responses. For both males
and females, the most frequent occupation response for real profiles was ``retired'',
a value which was extremely rare in scam profiles.

\begin{table}[t!]
\caption*{\textsc{Table I}: Topmost occupation areas by presented gender}
\centering
\scalebox{0.725}{
\setlength{\tabcolsep}{0.2cm}
\refstepcounter{table}
\subfloat[Male profiles]{\begin{tabular}{lrlr|}
  \hline
Real & Freq & Scam & Freq \\ 
  \hline
other & 0.15 & military & 0.25 \\ 
  self & 0.07 & engineer & 0.25 \\ 
  engineer & 0.07 & self & 0.10 \\ 
  tech. & 0.05 & business & 0.06 \\ 
  student & 0.05 & building & 0.06 \\ 
  retired & 0.05 & other & 0.04 \\ 
  building & 0.05 & contract & 0.04 \\ 
  service & 0.04 & medical & 0.03 \\ 
  transport & 0.04 & manager & 0.02 \\ 
  manual & 0.03 & sales & 0.02 \\ 
   \hline
\end{tabular}
\label{tab:occupation_male}}
%
%
\subfloat[Female profiles]{\begin{tabular}{|lrlr}
  \hline
Real & Freq & Scam & Freq \\ 
  \hline
other & 0.15 & student & 0.21 \\ 
  student & 0.10 & self & 0.16 \\ 
  carer & 0.08 & carer & 0.10 \\ 
  service & 0.06 & sales & 0.07 \\ 
  clerical & 0.06 & military & 0.05 \\ 
  teacher & 0.06 & fashion & 0.04 \\ 
  retired & 0.05 & business & 0.04 \\ 
  self & 0.04 & other & 0.04 \\ 
  medical & 0.04 & finance & 0.03 \\ 
  housewife & 0.03 & service & 0.03 \\ 
   \hline
\end{tabular}
\label{tab:occupation_female}}
}
\end{table}

Table~\ref{tab:occupation_male} presents a strong bias of scam profiles towards
military and engineering professions. The desirability of male military profiles is a
bias romance scammers are already well-known for
exploiting~\cite{whitty2013scammers}. The masculine and high-status image of
engineering might similarly explain its use by scammers. Other
professions listed display a similar approach: business (in many cases, the raw
response being ``businessman''), medicine (i.e., ``doctor'') and
contracting professions which might lend themselves to explanations for why a
person would later require money to be sent overseas.
As shown in Table~\ref{tab:occupation_female}, female scam profiles present less
clearly suspicious occupations, with `student' and `carer' groups leading. 
The appearance of `fashion' further down the list does speak towards a
desirability bias (e.g., ``model''). The `military'
group makes a surprising appearance---no real female profiles claimed such a
role---even more oddly, this
occupation is selected mostly by female profiles aged over 40. 
This may be an attempt to generalize the ``military scam'' used in male
profiles, but its strategy is unclear. 
For the most part, female scam
occupations fit with previous suggestions that scammers are exploiting the
desirability of a young, dependent female
partner in low-paying or non-professional
work~\cite{whitty2013scammers}. This role naturally lends
itself to an explanation for why a person might need financial support.

\subsubsection*{Location}

\begin{figure}[thb!]
\includegraphics[width=\columnwidth,trim = 0 0 0 0, clip]{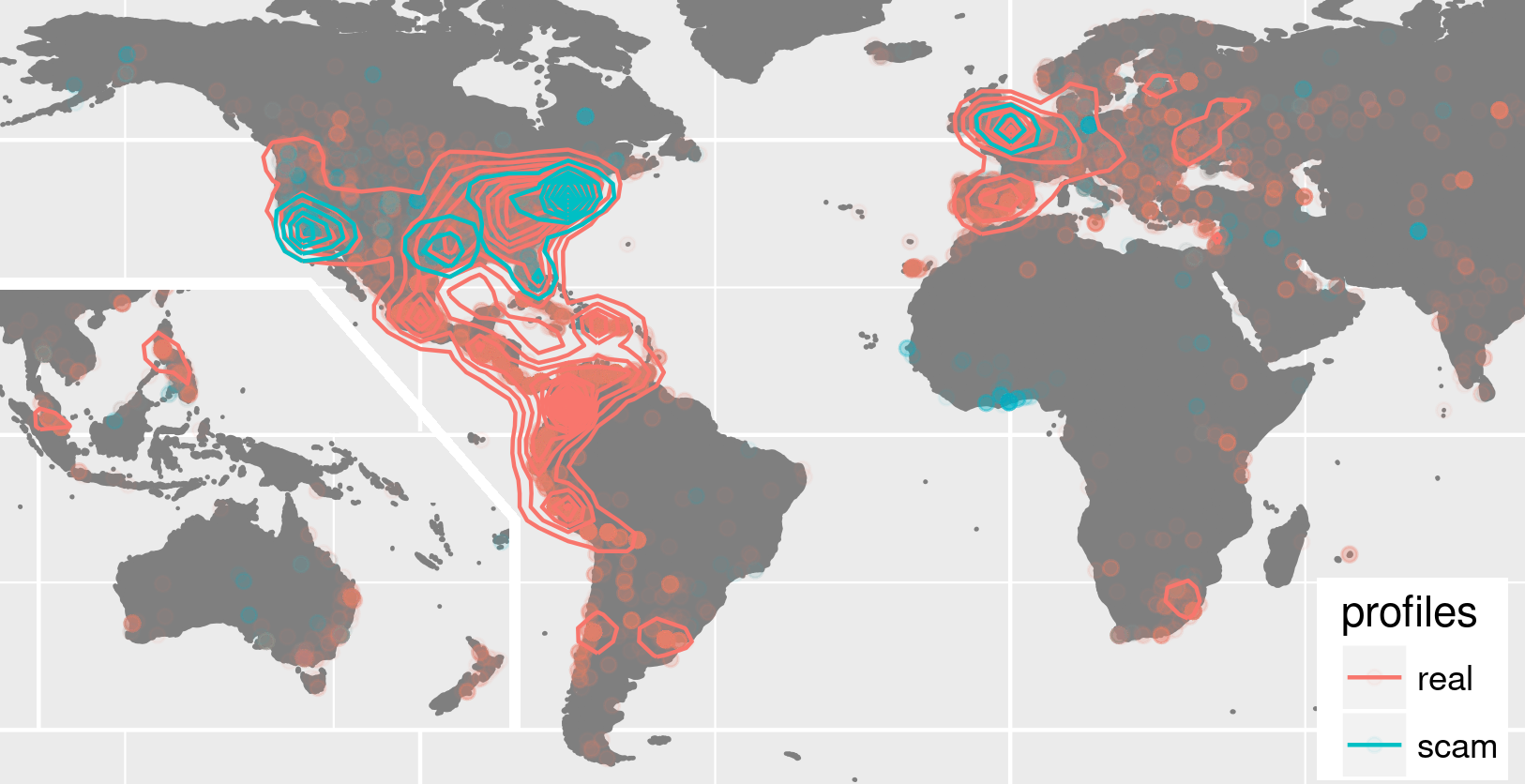}
\caption{Worldwide location of scammer and real profiles\gianluca{We should think about moving this figure to the appendix. Also, it is unreadable when printed in black and white}}
\label{fig:location}
\end{figure}

The location data reported in profiles was usually given to the city level,
although the specificity did vary, particularly within scammer profiles. The
original location responses were geocoded to provide lat/lon
points, and country of origin factors. 
As shown in Fig.~\ref{fig:location}, 
the scam profiles mostly claim to be in
the US or Western Europe. Corresponding with the earlier observation of a low
incidence of claimed Hispanic ethnicity, scammer profiles rarely claim to be
located in Latin America or Spain, despite a large real user population from these areas.
This suggests that a substantial
Spanish-speaking population of the dating site is not (yet) being targeted, 
possibly due to language barriers.


With regard to the targeted national locations, the
concentration of scammer profiles in the US is highly notable, with nearly three-quarters
of scam profiles with given locations claiming to be resident there. The
secondary targets were the UK and Germany. More plausibly honest
responses, such as Ghana, may be reactions to the dating site's 
methodology of comparing IP geolocations to declared location~\cite{edwards2018conPro}. The distribution 
suggests scammers are targeting rich,
Western and mostly English-speaking nations.

Scammer profiles most often declared locations which were well-known Western
cities. The most frequent city response was New York, being roughly 13\% of all
scammer locations, followed by Los Angeles (7\%) and then London, Dallas, Miami, Houston and Berlin. Selecting well-known and large cities 
avoids the need for intricate knowledge of a smaller city/town and makes it
easier for a scammer to remotely obtain enough detail to appear plausible. This
approach also enables a travel narrative---commonly, a wealthy businessman originally from a large city but
currently away on business.


\subsection{Image Recognition}


\begin{figure}[t!] 
\centering
\subfloat[Original.]{\includegraphics[width=.24\columnwidth]{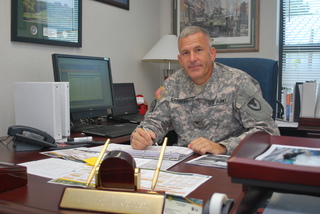}\label{fig:image-modification:A}}
\hfill
\subfloat[Fake Image.]{\includegraphics[width=.24\columnwidth]{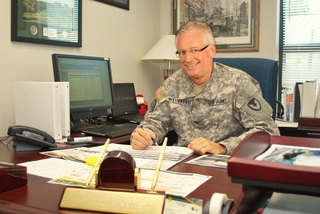}\label{fig:image-modification:B}}
\hfill
\hfill
\subfloat[Original.]{\includegraphics[width=.24\columnwidth, trim=1 3 2 3, clip]{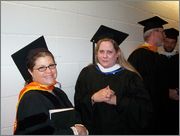}\label{fig:image-modification:4A}}
\hfill
\subfloat[Fake Image.]{\includegraphics[width=.24\columnwidth, trim=2 8.5 5 8.5, clip]{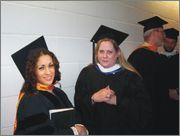}\label{fig:image-modification:4B}}
\\
\subfloat[Original.]{\includegraphics[width=.24\columnwidth, trim=1 20 2 4.5, clip]{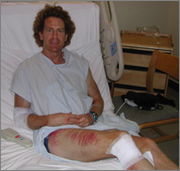}\label{fig:image-modification:3A}}
\hfill
\subfloat[Fake Image.]{\includegraphics[width=.24\columnwidth, trim=1 15 2 3, clip]{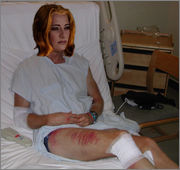}\label{fig:image-modification:3B}}
\hfill
\hfill
\subfloat[Original.]{\includegraphics[width=.24\columnwidth, trim=1 45 2 12, clip]{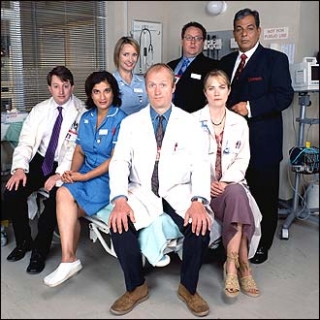}\label{fig:image-modification:2A}}
\hfill
\subfloat[Fake Image.]{\includegraphics[width=.24\columnwidth, trim=1 25 2 8, clip]{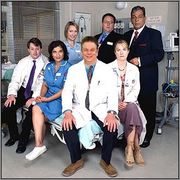}\label{fig:image-modification:2B}}
\caption{Certain contexts such as the military, academic or the medical one are often used to attract vulnerable users\protect\footnotemark.}\label{fig:image-modification}
\end{figure}
\footnotetext{Publicly available images from \href{https://www.romancescam.com/forum/viewtopic.php?f=13\&t=51587\#p289951}{https://www.romancescam.com/}}

The use of images plays an important role in online dating sites. The right set 
of pictures can both maximize the number of people interested in a profile and 
help users to limit interactions to a certain type of person. 
This can be leveraged by criminals to reach a larger number of potential victims 
and to attract vulnerable users. 

Scammers typically select the face of a publicly available image and 
build a fake persona with other images from different desirable contexts.  
The military context is recurrently exploited \cite{whitty2013scammers} 
as certain vulnerable victims seek a `knight in shining armor'.  
Other popular contexts are the academic and medical ones, where scammers 
pretend to be practitioners, students or patients. 
Fig.~\ref{fig:image-modification} shows how a scammer faked an image of a 
high-ranked military officer to impersonate a third party. 
We next study how to extract semantics from images to better understand the choices 
made by both legitimate users and scammers when selecting profile pictures.

As mentioned earlier, the data available in our dataset for the \textit{scammers} 
category and the \textit{real} category differs slightly. For the scammers dataset 
there are samples available with multiple images per profile. 
Conversely, for the real dataset there is 
usually only one picture per profile. Overall, 
we found images in approximately 65\% of the profiles\footnote{Without including
the default site avatar: 
\includegraphics[width=.5cm, trim = 0 10 0 0, clip]{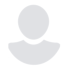}.}. 
While the proportion of profiles with images is equivalent in both types of 
categories, the absolute number of images per profile is larger in the scammers 
dataset. Specifically, there are 0.65 images per profile in the real dataset 
and 1.5 in the scammers one. 
Note that there may be variation in the distribution of images per profile category 
across different dating sites. However, in our dataset, fraudsters tend to share more information than legitimate users.

\begin{figure}[t] 
\centering
\subfloat[Real.]{\includegraphics[width=.24\columnwidth]{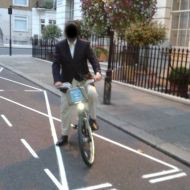}\label{fig:captions:A}}
\hfill
\subfloat[Real.]{\includegraphics[width=.24\columnwidth]{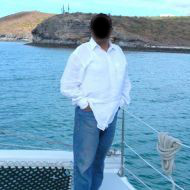}\label{fig:captions:B}}
\hfill
\hfill
\subfloat[Scammer.]{\includegraphics[width=.24\columnwidth]{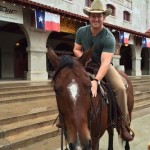}\label{fig:captions:C}}
\hfill
\subfloat[Scammer.]{\includegraphics[width=.24\columnwidth]{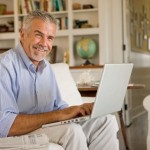}\label{fig:captions:D}}
\caption{Sample of images from our dataset. Faces from 
\textit{real} profiles have been redacted to preserve 
the anonymity.
}\label{fig:captions}
\end{figure}

A common theme found in both types of profiles is the use of pictures where users not 
only show their physical appearance, but also convey a sense of the hobbies or 
interests the person holds. Fig.~\ref{fig:captions} shows four examples of images 
found in our dataset where subjects are, for instance, riding or sailing.  

As it is relevant how scammers present themselves in their profile pictures, 
we next elaborate on how to extract meaningful information from them. 
Recent work in the field of computer vision~\cite{karpathy2015deep,vinyals2016show} 
has shown that it is possible to automatically describe the content of an image 
accurately using deep learning. 
The key idea is to train a deep network with a large corpus of 
images for which there is a ground truth of visible context.
The resulting network is then expected to i) know how to recognize elements 
appearing in the images and ii) be capable of generating an adequate description. 

For the purpose of this work, we mainly rely on~\cite{vinyals2016show} to build  
a generative model based on a deep Neural Network (NN). The system consists of a 
convolutional NN combined with a language-generating recurrent NN. 
The model has been built using a very extensive dataset distributed by Microsoft 
called COCO (Common Objects in Context)\footnote{\url{http://mscoco.org/}} 
with 
over 300,000 images. The output of the system is a meaningful description of the 
image given as input. 

For each image in a profile, we output the description that best represents
(according to the model) the semantics involved in the picture.
Fig.~\ref{fig:captions} shows images from four different profiles, two in the
\textit{real} category and two in the \textit{scammer} category. We next show
the full output extracted from the images shown in Fig.~\ref{fig:captions}.  For
each image we output three possible descriptions with probability $p$.


\vspace{.25cm}
\begin{center}
\noindent\fbox{\begin{minipage}{0.9\columnwidth}
\centering
{\footnotesize
Descriptions automatically generated from Fig.~\ref{fig:captions:A}:
\begin{enumerate}
\item A man riding a motorcycle down a street \hfill ($p = 72.2e-04$)
\item A man riding a bike down a street \hfill $(p = 29.3e-04)$
\item A man riding a bike down the street \hfill $(p = 3.7e-04)$
\end{enumerate}}
\end{minipage}}
\end{center}
\vspace{.1cm}

The descriptions shown above have been extracted from the image of a profile belonging 
to the \textit{real} category. The image shows a man standing over a bicycle in the 
street. The description in 1) guessed that the man is riding a motorcycle. This 
misconception can most likely be attributed to the headlight and the ad banner over it 
(uncommon in bikes). Descriptions 2) and 3) however are guessed correctly with
a probability of the same magnitude.  We argue that this type of mistake is
orthogonal to our problem. Confusing objects of similar kinds should not have a
negative impact as long as the main activity is correctly inferred (i.e.: a man
riding down the street).


\vspace{.25cm}
\begin{center}
\noindent\fbox{\begin{minipage}{0.9\columnwidth}
\centering
{\footnotesize
Descriptions automatically generated from Fig.~\ref{fig:captions:B}:
\begin{enumerate}
\item A man standing in a boat in the water \hfill $(p = 28.0e-05)$
\item A man standing in a boat in a body of water \hfill $(p = 9.9e-05)$
\item A man in a suit and tie standing in the water \hfill $(p = 2.1e-05)$
\end{enumerate}}
\end{minipage}}
\end{center}
\vspace{.1cm} 

The afore set of descriptions also belong to the \textit{real}
category. The image shows a man standing in the deck of a boat as correctly
predicted.


\vspace{.25cm}
\begin{center}
\noindent\fbox{\begin{minipage}{0.9\columnwidth}
\centering
{\footnotesize
Descriptions automatically generated from Fig.~\ref{fig:captions:C}:
\begin{enumerate}
\item A man riding on the back of a brown horse \hfill $(p = 11.8e-03)$
\item A man riding on the back of a horse  \hfill $(p = 1.3e-03)$
\item A man riding on the back of a white horse \hfill $(p = 0.9e-03)$
\end{enumerate}}
\end{minipage}}
\end{center}
\vspace{.1cm}

The descriptions shown above have been extracted from the image of a profile
belonging to the \textit{scammer} category. The image shows a young man riding a
brown horse.  All three descriptions complement each other by adding additional
details of the image.  It is common to find misappropriated images that do not
belong to the scammer---either because they have been stolen from a legitimate
profile or because they have been taken from elsewhere on the Internet). A
reverse search of the image does not reveal the source.


\vspace{.25cm}
\begin{center}
\noindent\fbox{\begin{minipage}{0.9\columnwidth}
\centering
{\footnotesize
Descriptions automatically generated from Fig.~\ref{fig:captions:D}:
\begin{enumerate}
\item A man sitting in front of a laptop computer \hfill $(p = 13.1e-03)$
\item A man sitting at a table with a laptop \hfill $(p = 3.6e-03)$
\item A man sitting at a table with a laptop computer \hfill $(p = 2.0e-03)$
\end{enumerate}}
\end{minipage}}
\end{center}
\vspace{.1cm} 

These descriptions belong to an image from the \textit{scammer} category. The
image shows a middle aged man sitting in front of a laptop and a table in the
background.  This image together with others found in the same profile are stock
images. 

It is worth noting the level of detail shown in each caption, which not only
identifies the main actor within the picture (a man in these cases), but also
the backdrop and the activity being undertaken.  


There are a number of common topics displayed across images in both profiles. 
When looking at the gender of the people present in the images, we 
can observe that males appear in about 60\% them as shown in Table
\ref{tab:typeImages}. This matches with the distribution of gender 
reported in the profile demographics. 

\begin{table}[h!]
\centering
\caption{Topics found across profiles with images.}
\scalebox{0.8}{
\begin{tabular}{lccc}
  \hline
 Type       &  Real Profiles & Scam Profiles & All  \\ 
\hline
  Male      &   57.75\%  &   63.76\%   &  60.48\% \\
  Groups    &   0.50\%  &    2.22\%  &  1.28\% \\
  Children  &  5.21\%   &   3.38\%   &  4.38\% \\
  Food      &  1.86\%  &  3.62\%  & 2.66\% \\
  Animals   &  0.77\%   &   1.08 \%  &  0.91\% \\
\hline
  Discriminant  &  13.76\% &  17.77\% &  15.58\% \\
\hline
\end{tabular}
}
\label{tab:typeImages}
\end{table}

There are also a number of topics slightly more prevalent in one or the other
profile categories, i.e.: group pictures (including couples), pictures with
children, or presence of food (e.g., wine, bbq, cake).  For instance, there are
over four times more \emph{group pictures} in the scammers category than in the
real one.  Contrastingly, the number of images with \emph{children} in real
profiles is almost double. 
Combining together all informative elements of the images, 
we can observe that about 15\% of images contain descriptions that 
appear exclusively in one of the two categories (referred to as 
`\emph{discriminant}' profiles in Table \ref{tab:typeImages}). 
This indicates that 
there is a large number of images for which their context can 
be used to characterize scammers. In other words, scammer profiles 
feature more pictures of certain groups. Note that fraudsters frequently
iterate through certain themes known to be appealing (e.g.: men in uniform).
This might also simply be down to the availability of images which the scammer
steals e.g., those from stock photo databases). The image shown in
Fig.~\ref{fig:captions:D}, for instance, is a stock image. 
\subsection{Profile Descriptions}
\label{sec:descriptions}

Contrary to most real-life encounters, on dating websites, a user can 
easily disclose very personal information, such as their life story, what they
are looking for in a partner, their hobbies, their favorite music, etc., to a 
complete stranger and without being interrupted. Moreover, filling in 
a personal description is usually highly encouraged by any dating
website, because it can capture other users' attention and
increase the chances of meeting a user's `perfect match'.

For scammers, however, the profile description provides yet another
means to mislead their victims. Prior research has shown 
that they will go to great lengths to create the `ideal' profile,
to gain a potential victim's interest and to maintain the pretense of a real (online) 
relationship~\cite{whitty2013scammers,yen2016case}. 
As a result, most scammers in our dataset---5,027 out of 5,402---attempted to create 
an attractive user account by advertising broad pretended interests and characteristics.
The real users were less inclined to provide such personal information about themselves: 
only 5,274 (out of 14,720) generated a profile description.

Recent advances in natural language processing technology have enabled researchers to perform automatic linguistic 
analyses of lexical, morphological, and syntactic properties of texts. However, most traditional studies use large 
sizes of training data with a limited set of authors/users and topics, which usually leads to a better performance of 
the machine learning algorithms. Profile descriptions are, however, typically short and can include a 
whole range of different topics. With regard to the dataset described in this paper, the average number of words 
per profile description was 78.7, with scammers producing more words on average (104.5) than genuine users (54.1),
This effect is so pronounced that despite there being fewer scammers than real users, the overall total of 525,336 words for the \textit{scam} category was greater than the 285,407 words for the \textit{real} category. 
The finding that scammers' profiles have a higher word count compared to genuine
profiles is consistent with previous literature stating that liars tend to
produce more words~\cite{hancock2005automated}. 
To analyze the variety of topics that are present in our dataset, we used dictionary terms that are mapped to 
categories from the LIWC 2015 dictionary~\cite{pennebaker2015development}. Category frequencies were recorded for 
each 
profile description. Our results showed that scammers referred considerably
more to emotions---both positive and negative---than genuine users.
Additionally, they use words related to family, friendship, certainty, males
and females more often, while real users tend to focus on their motives or
drivers (e.g., affiliation, achievement, status, goals), work, leisure, money,
time and space. With regard to language use, we found that scammers use more
formal language forms, while genuine users displayed more informal language
forms (e.g., Netspeak).  



\section{Classifying False Profiles}
\label{sec:core}

\begin{figure*}[t]
\centering
\includegraphics[width=0.9\textwidth, trim = 20 195 15 35, clip]{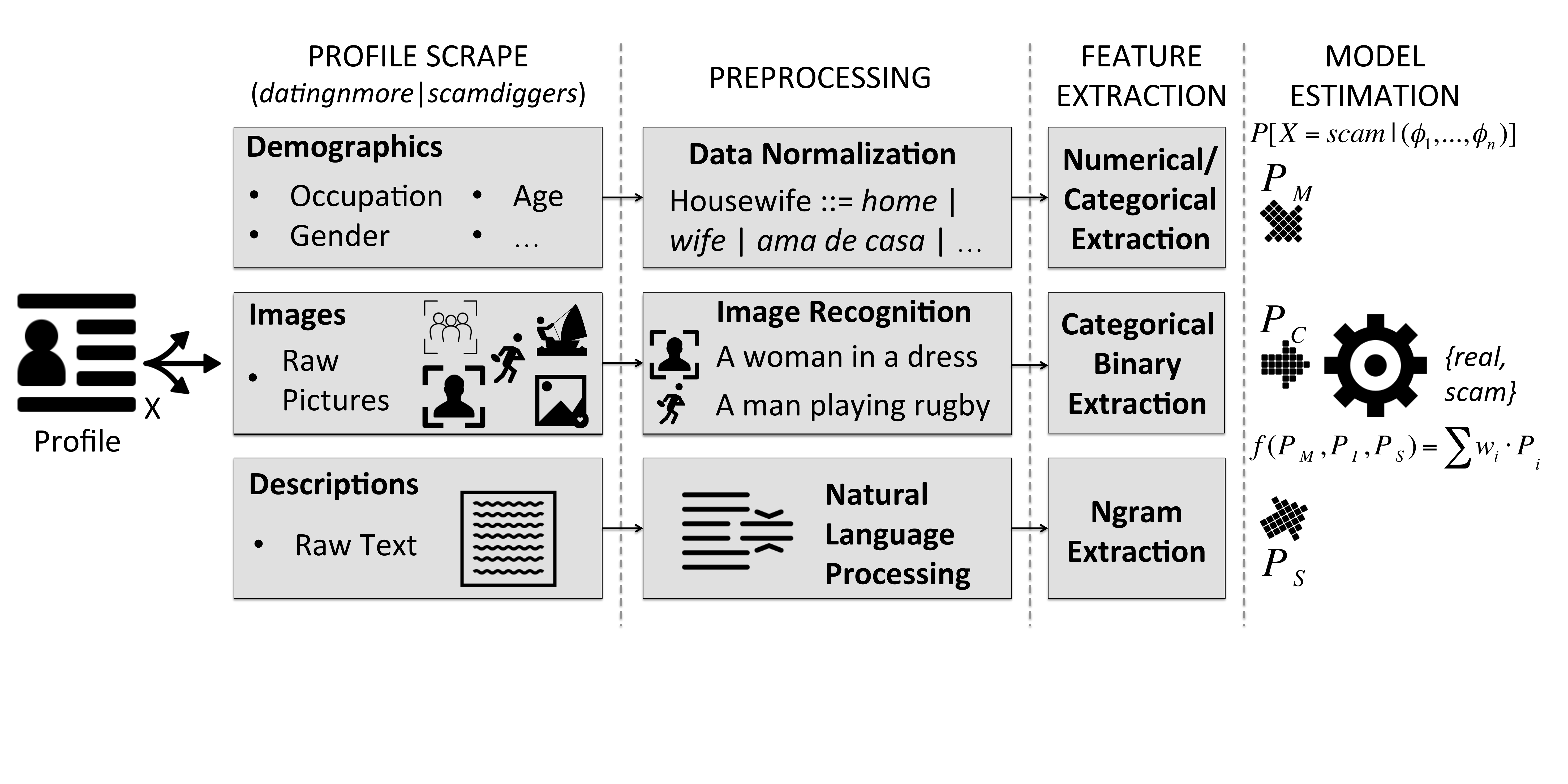}
\caption{Key features extracted from dating profiles.}\label{fig:architecture-general}
\end{figure*}


\label{sec:core:overview}

%
A high-level overview of our system can be obtained from Fig.~\ref{fig:architecture-general}. 
The system is first trained using a dataset of real and scam profiles. The goal of 
this phase is to obtain the following key elements that will later be used to 
identify fraudsters:

\begin{enumerate}
\item[(i)] A set of prediction models $\mathcal{P} = \set{P_1, \ldots, P_i}$ that output the probability 
$$\theta_i(\phi_1, \ldots, \phi_n) = P_i[X=\texttt{scam}\mid(\phi_1, \ldots, \phi_n)]$$
of each profile $X$ being \texttt{scam} given a feature vector $(\phi_1, \ldots, \phi_n)$ obtained from different profile sections $i$. 

\item[(ii)] A weighted model 
$f(\mathcal{P}) = \sum w_i \cdot P_i $
that combines all individual predictions in $\mathcal{P}$. Here, each individual 
classifier $P_i$ is weighted by $w_i$ according to the accuracy given on a validation 
set that is different from the training one. This will also serve as a way to 
calibrate individual probabilities. The final classifier will then output a decision 
based on a vote such that 
$$
f =
\begin{cases}
\texttt{scam} & \text{if~} f(\mathcal{P}) < \tau\\
\texttt{real} & \text{otherwise,}
\end{cases}
$$
where $\tau$ is a threshold typically set to $\floor*{\frac{\sum w_i}{2}} + 1$.
\end{enumerate} 

For the sake of simplicity, we refer to the model presented in (ii) as 
\emph{weighted-vote}. One can simplify the model by giving equal weight 
to all $w_i$ (typically $w_i = 1$) and obtaining a nominal value for $P_i$
before voting. In other words, applying a threshold for each $P_i$ (e.g.,  
$0.5$) and creating an equal vote among participants. We refer to this 
non-weighted voting system as \emph{simple-vote}. 



\subsection{Feature Engineering}
\label{sec:core:fengineering}

Our work considers a diverse set of features in order to build a robust 
classification system. The proposed set of features contains elements obtained 
from three different sources: 
(i) structured attributes of the profile referred to as \emph{demographics} 
and denoted as $\theta_M$, 
(ii) features extracted from raw images referred to as \emph{captions} ($\theta_C$), and 
(iii) features extracted from unstructured text (\emph{description} 
denoted as $\theta_S$). 
Based on the preprocessing described in Section~\ref{sec:preliminaries}, we
extract different types of features as described below. 

\begin{itemize}
\item \textbf{Numerical Features} (NF): refers to those attributes from a profile 
that take a quantitative measurement such as the age (18-85) of a person.  
\item \textbf{Categorical Features} (CF): refers to those attributes 
that take a limited number of possible values such as the gender (male or female). 
\item \textbf{Set-based Features} (SBF): refers to those attributes that can take an 
arbitrary number of values and the relationship between sets of attributes is 
relevant (e.g., words in the description). 
\end{itemize}



Table \ref{tab:features} shows the set of features proposed categorized by the 
source in the profile. 
For $\theta_M$ we considered the age, and the 
Cartesian location values as numerical values, while other attributes were
treated as categorical. As described previously, common occupation responses were grouped
into 45 different occupation areas (e.g., self-employed, military, legal). 
Long-tail occupation responses outside of these categories were grouped under
\emph{other}. In training, no such response appeared more than twice, and 
85\% of real and 96\% of scammer occupations were captured by known
categories.

\begin{table}[t]
\centering
\caption{Our proposed set of features.}
\begin{tabular}{lllll}
  \hline
 ID & Source & 	Name 		& Type 	& $|\theta_i|$	\\ 
\hline
\multirow{7}{*}{$\theta_M$} &
\multirow{7}{*}{Demographics}
		  	 		  	& 	Age 		& 	NF 	& 	\\ 
		  	& 		  	&  	Gender 		& 	CF & 	 	\\ 
		  	& 		  	&  	Latitude 	& 	NF 	& 	\\ 
		  	& 		  	& 	Longitude 	& 	NF 	& 	\\ 
		  	& 		  	& 	Country 	&	CF & 	237 \\ 
		  	& 		  	& 	Ethnicity	& 	CF & 	\\ 
		  	& 		  	& 	Occupation 	& 	CF & 	\\ 
		  	& 		  	& 	Marital Status 	& 	CF & 	\\ 
\hline
\multirow{3}{*}{$\theta_C$} &
\multirow{3}{*}{Captions}
		  	 		  	&	\texttt{set}(\emph{entities}) 		& 	CF & \multirow{3}{*}{363}	\\ 
		  	& 		  	&	\texttt{set}(\emph{actions}) 		& 	CF & 	\\ 
		  	& 		  	&	\texttt{set}(\emph{modifiers}) 		& 	CF & 	\\ 
\hline
$\theta_S$ 	&
 Descriptions & \texttt{set}(\emph{ngrams}) 	& 	SBF & 105,893	\\ 
\hline
\end{tabular}
\label{tab:features}
\end{table}

For $\theta_C$, after extracting the most representative caption per 
image, we removed the least informative elements and retained only 
entities (nouns), actions (verbs), and modifiers (adverbs and adjectives). 
Each element in the caption was stemmed to a common base to reduce 
inflectional forms and derived forms. 
Rather than treating the captions as set-based features, we encoded them 
as categorical features, as the order of the actions 
is not relevant in this document vector approach. Generated captions are simple and their structure 
always follows the same pattern. The presence of a given action and 
the set of objects appearing in the image is itself informative.  
Encoding the relationship between the different parts of speech would 
unnecessarily increase the number of features. 

Finally, we extracted set-based features from the textual content of the 
tokenized descriptions ($\theta_S$).
We considered Bag-of-Words (BOW), word \emph{n-}grams, 
character \emph{n-}grams and LIWC features. 
Because
word bigrams (\emph{n-}grams of length $n=2$) yielded the best results during our 
preliminary experiments and combining different feature types did not lead 
to better classification results, we only included
word bigrams in the rest of the experiments. Additionally, 
stemming and stop word removal resulted in a worse performance, so all
features were included in their original form during the experiments.


\subsection{Prediction Models}
\label{sec:core:model}

Because most of the fields are optional, user profiles in online 
dating sites are inherently incomplete. Some users are uncomfortable with high levels of 
self-disclosure and some are more interested in contacting others than presenting details about
themselves~\cite{whitty2008revealing}. Thus, any reliable 
detection system should be able to flexibly deal with incomplete profiles. In this section,
we present three independent classifiers to estimate the presence of 
fraudulent profiles. Each classifier is designed to effectively model 
a section of the profile (based on $\theta_M$, $\theta_C$ and $\theta_S$ as 
described previously). Probability outputs from each classifier are later 
combined to provide one balanced judgement. 
By using multiple classifiers designed on individual sections of the profile, we increase
the likelihood that at least one classifier is capable of making an informed
decision.
Moreover, ensembles often perform better than 
single classifiers~\cite{dietterich2000ensemble}. 


\subsubsection*{Demographics Classifier}

The demographics classifier uses the greatest variety of original profile
attributes. Unlike the image and description classifiers, handling this feature
set means dealing with non-binary missing data situations---location and
ethnicity might be missing for a given profile which still contains age and
gender information. When no data is available, the least-informative prior
is the base rate of real vs scam profiles, as is used in the image
and description classifiers. In most situations within the demographics
data more information than this is available, and should be used.

Approaches for handling missing data---in the situation where the case cannot be
discarded, such as appearing in a test or validation set---reduce to either
some form of imputation or the use of a classifier robust to missing values,
such as a Naive Bayes approach. Given problematic randomness assumptions
inherent to the most useful imputation methods, we opt to use a Naive Bayesian
classifier to handle prediction for profiles with missing data attributes.

However, Naive Bayes is not the most effective classifier for profiles with all
data present---a significant proportion of the dataset. For this subset, it is 
more effective to use a classifier which performs better and does not handle
cases with missing data. In our case, a Random Forests model was selected. The
final approach to providing $P_M(X = \texttt{scam})$ is to train a joint
Random Forests and Naive Bayes model, using the high-performing Random Forests 
model to make predictions where all demographic data is available, and the
gracefully-degrading Naive Bayes model for all other cases.

\subsubsection*{Images Classifier}

We build a prediction model based on the features extracted from the
captions of the images such that $P_C(X = \texttt{scam})$. The architecture 
of our system is highly flexible and accepts a wide range of classifiers. 
Our current implementation supports Support Vector Machine (SVM), 
Random Forests and Extra Randomized Trees. 
For the purpose of this paper, we selected SVM with radial kernel as the 
base classifier for the images. 

SVM has been successfully 
applied to fraud detection in the past~\cite{bhattacharyya2011data} and 
has been shown to have better performance compared  
to 179 classifiers (out of 180) on various datasets~\cite{Fernandez-Delgado:2014:WNH}. 
SVM also tends to perform well when the number of samples is much greater 
than the number of features, as it is the case here. 
In addition, it is 
less sensitive to data outliers---instead of minimizing the local error, 
SVM tries to reduce the upper bound on the generalization error. 


\subsubsection*{Descriptions Classifier}


Previous work~\cite{yen2016case} has shown that scammers attempt to keep their labor costs down to be able to exploit  
different social media and to continuously produce the interaction that is required 
to make their on-going scams succeed. To achieve this goal, they tend to edit pre-written 
scripts that are often shared on underground forums---labeled by the ethnicity,
age group, location and gender of the potential victim. Hence, for providing $P_S(X = \texttt{scam})$, we compared the performance of two approaches: (\textit{i}) a similarity-based approach, in which we applied shingling ($k = 5$) to extract the set of all substrings from each profile description in training and calculated the Jaccard similarity for each pair of profile descriptions (see ~\cite{leskovec2014mining}); and (\textit{ii}), we trained an SVM algorithm 
(linear kernel) as implemented in LibShortText~\cite{yu2013libshorttext},
an open-source software package for short-text classification
and analysis. Parameters for both approaches were experimentally determined on
a small subset of each training partition during cross
validation. Within the SVM experiments, features were represented by TF-IDF scores, which reflect
the importance of each feature (in this case, each word bigram) to a document 
(i.e. a user profile description) in terms of a numerical frequency statistic over
the corpus~\cite{leskovec2014mining}.
\matthew{Reworded}

\subsubsection*{Ensemble Classifier}
%
The goal of this method is to combine the predictions of the base estimators 
described above to improve the robustness of the classification. Ensemble 
methods are designed to construct a decision based on a set of classifiers by 
taking a weighted vote of all available predictions. In our system, we have a 
function $f$ that is estimated using an independent set of samples. This 
function will then be used during testing to weight each prediction model 
$P_i$ such that: 
$f(P_M, P_C, P_S) = \{\texttt{scam}, \texttt{real}\}.$

For the decision function $f$ we use a Radial Basis Function (RBF) that 
measures the distance to the center of the SVM hyperplane bounding each 
$P_i$. This function is defined on a Euclidean space and it only measures 
the norm between that point and the center (without considering the angular 
momentum). This function is approximated with the following form
$$f = \sum w_i \delta(||p_i||),$$
which can be interpreted as the sum of the weights $w_i$ times the probability 
score $p_i \in P_i$ given by the individual classifiers in the voting system 
described above. 

\subsubsection*{Single Classifier}
%
We compare the results of our ensemble method to the predictions made by a single 
SVM classifier (linear kernel) in which all demographics, captions and description 
features are included in each document instance. Features were represented by their 
absolute values and parameters were again experimentally determined on
a small subset of each training partition during cross
validation.


%


\section{Evaluation}
\label{sec:evaluation}


\label{sec:evaluation:setup}

In evaluating and developing the classification system described 
before, 
we applied the following methodology.

\subsubsection*{Methodology}

We divided the dataset into a 60\% training set, a 20\% test set
and a 20\% validation set. Profiles were assigned to each set
randomly under a constraint preventing variants of the same scam 
profile from being assigned different sets or folds. Development of the
classification system proceeded as follows:

\begin{enumerate}
	\item Each component classifier was designed within the 60\% training set, and
individual performance levels established through ten-fold cross-validation within 
this set. 
	\item Once classifier design was complete, each component classifier was trained
on the full training set.
	\item To design the ensemble model, each classifier produced probabilities
and labels for the test set. The ensemble was developed on these probabilities,
and performance was established through five-fold cross-validation on the test set.
	\item Based on performance within the test set, the ensemble model and
the choice of outcomes to report within the validation set was decided.
	\item For final validation, individual classifiers were trained on the
training set, produced probabilities and labels for the testing set and the
validation set, the ensemble model was trained on the probabilities given for
the test set, and its predictions taken for the validation set. 
	\item The single classifier was trained on the combination of the training 
	and test data and evaluated on the validation set.
\end{enumerate}

\subsection{Classification Results}
\label{sec:evaluation:results}

We present our results together with a number of case studies, covering 
all four dimensions of the classification performance: 
(i) scam profiles correctly classified (TP), 
(ii) real profiles correctly classified (TN), 
(iii) real profiles misclassified (FP), and
(iv) scam profiles misclassified (FN).

\subsubsection*{Summary}

\begin{table}[t]
\setlength{\tabcolsep}{0.09cm}
\centering
\caption{Final results for each component classifier, simple majority
voting, a similarity-only approach, a single classifier using all features, and the weighted-vote ensemble}
\scalebox{0.8}{
\begin{tabular}{lrrrrrrrr}
  \hline
\textsc{classifier} & \textsc{tn} & \textsc{fn} & \textsc{fp} & \textsc{tp} & \textsc{prec.} & \textsc{rec.} & \textsc{f1} & \textsc{acc} \\ 
  \hline
demographics & 2725 & 196 & 149 & 903 & 0.858 & 0.822 & 0.840 & 0.913 \\ 
  captions & 2872 & 499 & 2 & 600 & 0.997 & 0.546 & 0.705 & 0.874 \\ 
  description & 2758 & 215 & 116 & 884 & 0.884 & 0.804 & 0.842 & 0.917 \\ 
  similarity-only & 2939 & 435 & 28 & 571 & 0.953 & 0.568 & 0.712 & 0.884 \\
  simple-vote & 2870 & 189 & 4 & 910 & 0.996 & 0.828 & 0.904 & 0.951 \\ 
  single & 2820 & 108 & 54 & 1027 & 0.950 & 0.905 & 0.927 & 0.959 \\
  weighted-vote & 2834 & 78 & 40 & 1021 & 0.962 & 0.929 & 0.945 & 0.970 \\ 
   \hline
	Excluding new variants & \\
   \hline
 demographics & 2725 & 122 & 149 & 569 & 0.792 & 0.823 & 0.808 & 0.924 \\ 
  captions & 2872 & 378 & 2 & 313 & 0.994 & 0.453 & 0.622 & 0.893 \\ 
  description & 2758 & 119 & 116 & 572 & 0.831 & 0.828 & 0.830 & 0.934 \\ 
  simple-vote & 2870 & 129 & 4 & 562 & 0.993 & 0.813 & 0.894 & 0.963 \\ 
  weighted-vote & 2818 & 53 & 56 & 638 & 0.919 & 0.923 & 0.921 & 0.969 \\ 
  \hline
	Excluding all variants & \\
  \hline
 demographics & 2707 & 114 & 167 & 577 & 0.776 & 0.835 & 0.804 & 0.921 \\ 
  captions & 2874 & 426 & 0 & 265 & 1.000 & 0.384 & 0.554 & 0.881 \\ 
  description & 2731 & 171 & 143 & 520 & 0.784 & 0.753 & 0.768 & 0.912 \\
  simple-vote & 2860 & 159 & 14 & 532 & 0.974 & 0.770 & 0.860 & 0.951 \\ 
  single & 2829 & 98 & 45 & 592 & 0.929 & 0.858 & 0.892 & 0.960 \\
  weighted-vote & 2841 & 69 & 33 & 622 & 0.950 & 0.900 & 0.924 & 0.971 \\ 
   \hline
\end{tabular}
}
\label{tab:results}
\end{table}

\begin{figure}
\centering
\includegraphics[width=.9\columnwidth, trim=0 15 30 20, clip]{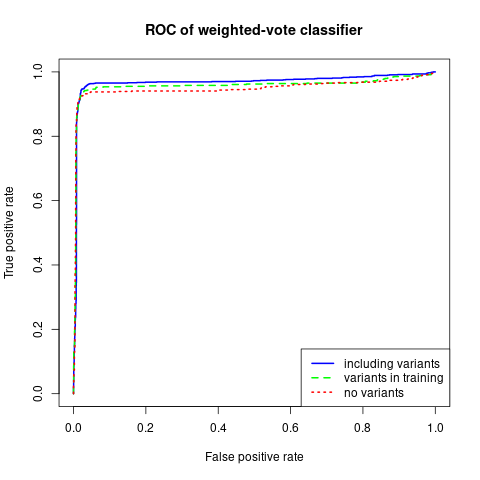}
\caption{ROC for the ensemble classifier}
\label{fig:ROC}
\end{figure}

Table~\ref{tab:results} presents the results within the validation set for each
classifier, for simple majority voting between all three classifier outputs, and 
for the SVM ensemble model trained on the classifier probabilities given for the
test set. Precision, recall and F1 are given for predicting scam profiles (the
minority class). 
Judging performance by F1, the best individual classifier was the SVM description
classifier (F1 = 0.842). As can be expected, the similarity-based approach (threshold 
Jaccard similarity of 0.259) yielded a high precision score, but a low recall 
score, which resulted in a markedly lower F1 of 0.712.
The demographics classifier was the next best component classifier (F1 = 0.840), but the captions
classifier was highly precise, making only two false-positive judgements.
Simple majority voting between classifier labels improved performance
significantly compared to any individual classifier, raising F1 to 0.904,
with a precision of 0.996. A single classifier using all features outperformed majority voting (F1 = 0.927).
The ensemble system outperformed both the single classifier and majority voting at 0.945 F1, 
significantly improving recall whilst maintaining a high level of precision. Over 97\% of all
profiles were classified correctly. Fig.~\ref{fig:ROC} characterises the ROC
performance for this ensemble depending on whether variants (near-duplicate profiles) were
excluded.

\subsubsection*{Feature Analysis}

We describe some of the most important features as identified by our classifiers.

Table~\ref{tab:rf_ranking} presents the total decrease in node impurities from
splitting on the each feature in the RF component of the demographics model, 
averaged over all trees. The most important feature was the occupation area reported 
in the profile. Node purity rankings are known to bias towards factors with many levels,
but the size of the interval between the occupation area and the other features suggests
that this ranking is genuine. This would accord with our observations in~\ref{sec:demographics}
about the use of occupation area as an attractive status marker. 

Table~\ref{tab:top_bigrams} presents the highest-weighted bigrams from the descriptions classifier
for the purpose of predicting the scam category. The most informative features tend to relate to nonfluencies
in English (starting descriptions with `Im' or `Am', constructs like `by name')
and attempts to overtly signal a romantic or trustworthy nature
(e.g.,``caring'', ``passionate'', ``loving''). Our topic analysis in
Section~\ref{sec:descriptions} also captures this tendency of scam profiles to
include more emotive language.

Table~\ref{tab:top_captions} presents the most discriminant features for the captions classifier. 
Features with a negative weight are more informative when discriminating real profiles. 
Instead, features with positive weight relate to scam profiles. 
Interestingly, some of the top elements embedded in the images map with relevant traits observed in the demographics such as the occupation (e.g., military) or the gender (e.g., male) c.f. \S\ref{sec:demographics}.

\begin{table}[ht]
\caption{Top-ranked features for component classifiers}
\centering
\scalebox{0.85}{
\subfloat[Feature ranking demograpics RF]{\begin{tabular}{rr}
  \hline
feature & purity \\ 
  \hline
occupation & 332.70 \\ 
  latitude & 198.02 \\ 
  status & 128.76 \\ 
  longitude & 128.71 \\ 
  age & 114.24 \\ 
  ethnicity & 110.53 \\ 
  gender & 64.52 \\ 
   \hline
\end{tabular}
\label{tab:rf_ranking}}
\subfloat[Top-weighted bigrams for scam descriptions]{\begin{tabular}{rr}
\hline
bigram & weight \\
  \hline
\emph{$<$start$>$} im & 0.3086\\
don t & 0.2318\\
caring and & 0.1776\\
and caring & 0.1674\\
by name & 0.1644\\
\emph{$<$start$>$} am & 0.1643\\
am just & 0.1641\\
that will & 0.1572\\
am here & 0.1568\\
tell you & 0.1481\\
\hline
\label{tab:top_bigrams}
\end{tabular}}
\subfloat[Feature raking captions]{\begin{tabular}{rr}
\hline
Keyword & weight \\
\hline
pizza & -1.0 \\
picture & -0.52 \\
child & -0.50 \\
bottle & -0.46 \\
christmas & -0.46 \\
\hline
driving & 1.0 \\
military & 1.0 \\
birthday & 2.0 \\
group & 2.46 \\
male & 2.95 \\
\hline
\label{tab:top_captions}
\end{tabular}}
}
\end{table}



\subsubsection*{True Positives}


About 98\% of the scam profiles have been detected by at least 
one of the classifiers. Consensus between classifiers accounts for the
majority of TPs, but performance improves yet further when
resolving disputes using classifier weights learned 
on an independent sample in the ensemble voting scheme.
Under this scheme, we manage to detect about 93\% of the fraudulent profiles with 
a high degree of confidence, compared to
81\% when only relying on the simple voting scheme. Roughly 
36\% of scammers were identified by all three classifiers.

For illustration, we present one TP case randomly chosen from those 
identified with a high degree of confidence. 
This is the case of a profile presenting as a 26 year old African American 
female, with the occupation reported as \textit{``studant''}. In the 
description, we can see certain traits uncommon in legitimate profiles, like the intention to establish a \textit{``great friendship''}. 
The misprint on the occupation and the poor language proficiency 
in the description might indicate that the fraudster 
lacks fluency in the English language.


\begin{quote}
\footnotesize
Hi.,
Am Vivia. How are you doing? I would be very happy to have a great friendship with you. my personal email is ( $<$user$>$@hotmail.com ) I look forward to hearing from you,
Vivia
\end{quote}


There are two images in the profile, which match the demographics reported. 
Specifically, one of the images has a young woman sitting on a park bench, 
while the other shows the same woman sitting in a study room with a 
laptop. The prevalence of certain elements such as the use of laptops 
across stock scam profile images could explain the decision taken by the image
classifier. 
%

\subsubsection*{True Negatives}

All real profiles have been identified as such by at least one of the classifiers. 
When combining the decisions, our system correctly classified 99.9\% real 
profiles using \emph{simple-vote} and about 98.6\% using \emph{weighted-vote}. 

Our randomly selected exemplar case is that 
of a 55 year old white woman. 
This user is based in Mexico, a location comparatively underpopulated by scam
profiles. It is worth noting that the age also deviates significantly 
from the average age of female scammers (30). 

This profile had an avatar image showing the face and shoulders of a woman. 
All the elements in the image looked conventional,
which is most likely why the classifier identified the profile as real. It is 
worth noting the poor quality of the image. Although the 
quality of the images was not measured in this work, we noticed that fraudsters care 
about it, and intend to investigate this further.
%
Comparing the profile description to the previous TP example, the overall
fluency is notably higher, both in terms of English grammar and the appropriate
format (as a self-description rather than a message). The user describes herself
and her interests, rather than focusing entirely on the reader.
\subsubsection*{False Positives}


There are a total of 4 real profiles misclassified 
under our most precise setting, for a false positive rate of 0.1\%. For those 4
profiles there was at least one classifier that 
correctly predicted the profile and, interestingly, all three classifications 
were available. This means that none of the errors can be attributed to a 
lack of information. Under weighted-voting the false positive rate 
rises to 1.4\%. When looking at the errors common to both voting 
schemes, we only find 2 misclassified real profiles. Both errors 
come from predictions by the demographics and descriptions classifiers.

From the demographics, we see that one profile is widowed, common amongst
scammers, and the other is mixed-race, which is slightly more common amongst
scammers. Both users claim locations with high scammer population: Texas
in the US, and close to London in the UK.
One of the profile descriptions was very short, and focused on the nature of their
intended relationship and the qualities of the intended partner. The other was
long, but both referred to topics which are more strongly correlated with
fraud, such as relationships and positive emotions.

\subsubsection*{False Negatives}

For misclassified scam profiles, weighted-voting produced 78 errors 
and simple-voting produced 189 errors. Out of all errors, we find 22 
cases where all three individual classifiers failed. Manual analysis of these cases
reveals that certain parts of the profiles 
look genuinely normal. In general, the image caption classifier was most likely to produce false
negatives, and overall errors occurred when either of the demographics
or description classifiers agreed. Of the two, the description
classifier was slightly more likely to produce FNs.

Such is the case of a 45 year old American solider from 
Louisiana named Larry. While the description is ordinary, the occupation 
raises suspicions due to the prevalence of military scammers. 
When there is a technical draw between the 
demographics and the descriptions and the images are not informative enough 
one can only aim at extracting additional features. 
For instance, 
in the case of Larry, the messages he exchanged with a victim are
valuable. The excerpt below shows a message sent to one of his victims 
wherein one may observe distinguishable wording and a clear manipulative
strategy in parts highlighted below: 
\begingroup
\addtolength\leftmargini{-0.2in}
\begin{quote}
\footnotesize
I must confess to you, you look charming and from all I read on your profile \emph{Id want you to be my one special woman}. I wish to build a one \emph{big happy family} around you. Im widowed with two girls, Emily and Mary, I lost their mom some years ago and since then \emph{Ive been celibate}. \emph{I think youve got all it takes to fill the vacuum left by my late wife to me and the kids}. I seek to grow old with you, children everywhere and grey hairs on our head. \emph{I wanna love you for a life time}. Hope \emph{to read from} you soon. You can addme on facebook $<$user$>$ or mail me at $<$user$>$@yahoo.com. Looking forward to your message. Regards, Larry.
\signed{A message sent to a victim}
\end{quote}
\endgroup

Adding such elements from later steps in the scam life-cycle 
could help to cover the cases where all three individual classifiers failed. 
However, these features are not necessarily available for all deployment
situations, and would restrict more general application of the model.

\section{Discussion}
\label{sec:evaluation:discus}


In this paper, we established the need for a systematic approach that 
automates online romance scammer detection. The system we presented 
is a key first step in developing automated tools that are able
to assist both dating site administrators and end-users in identifying
fraudsters \textit{before} they can cause any harm to their victims.
However, there are risks in deploying
such automated systems, principally:

 
\begin{enumerate}[noitemsep]
\item[i)]  Risk of denying the service to legitimate users.
\item[ii)] Risk of having scammers that have evaded the system. 
\end{enumerate}

\noindent We next discuss these implications and the limitations.

\subsection{Context-based Performance Maximization}

\guillermo{A bit of cutting potential here after adding previous subsection}

Under our current system, 96\% of profiles identified as scammers truly are, and
about 93\% of all scammers are detected. This performance is optimized for the
harmonic mean of these rates. One might further tune the model, either for 
minimizing the detection of \emph{false positives} or minimizing the \emph{false negatives}. 
This decision will rely on the priorities of the user of a classification tool:

\begin{description}
\item[Minimizing FP] -- when real profiles are misclassified, users are
inconvenienced by being flagged as scammers and are likely to be annoyed at a 
platform that does this. Thus, detection systems being run by dating sites must review
alerts or risk losing customers. To reduce workload and costs, dating
sites may want to minimize the risk of misclassifying real users, and use
user-reporting and education tools to catch scammers who evade preemptive detection.

\item[Minimizing FN] -- when scam profiles are misclassified, a user risks being
exposed to a scammer and suffering emotionally and financially as a result.
Given that the opportunity cost is comparatively low for potential partners being filtered out, a
``better safe than sorry'' attitude is justified. As such, safe-browsing tools 
that a user deploys themselves may wish to bias towards always flagging scammer
profiles, as the user may always disable or ignore such a tool if convinced it
is in error\footnote{Such tools may of course also allow the user to define their
own risk tolerances.}.
\end{description}

The simple voting classifier presented in Table~\ref{tab:results} provides an
easy example of a system biased towards minimising false positives, with only 4
appearing in a set of nearly 3,000 real profiles. If all three classifiers were
required to agree before a profile was classified as a scam (unanimous voting), then the false
positive rate would be too low for this study to detect (0 observed). 
Alternately, if the firing of any classifier was sufficient reason to
flag a profile, only 22 scammer profiles in over 1,000 would escape being
flagged.

Returning to the better-performing machine-weighted voting system, the ensemble
system could be optimized for any risk ratio by optimization towards a modified
F-score. The F-score can be weighted towards any desired ratio of
\emph{precision} (minimizing false positives) and \emph{recall} (minimizing
false negatives) by adjusting the $\beta$ parameter in the general equation:

$$
  \centering
  \displaystyle
  F_\beta = (1 + \beta^2) \cdot \frac{\mathrm{precision} \cdot \mathrm{recall}}{(\beta^2 \cdot \mathrm{precision}) + \mathrm{recall}}
$$

where the $\beta$ expresses the ratio by which to value recall higher than
precision. By selecting an appropriate balance between these measures and
then evaluating classifiers against this measure, a weighted voting system can
be tuned to individual risk tolerances.

\subsection{Comparison with Moderator Justifications}
\label{appendix:justifications}

The moderators who identified profiles as romance scammers provide a list of
justifications for their decision on each profile. By analyzing the given justifications, 
we can examine our classifier's performance next to individual human strategies for
scammer identification.

Table~\ref{tab:justifications} presents figures for the proportion of scam
profiles labeled with common justifications. The figures are counted in terms
of profiles and not profile-variants. Alongside figures for all scam profiles
are figures for the validation set upon which the ensemble system was tested,
and figures for the scam profiles which the ensemble classifier mislabeled as
non-scam profiles (false negatives). 

\begin{table}[t]
\setlength{\tabcolsep}{0.1cm}
\centering
\caption{Comparison of overall, validation and false-negative incidence of
moderator justifications for scam-classified profiles}
\scalebox{0.85}{
\begin{tabular}{lrrrrrrr}
  \hline
\textsc{Reason} & \textsc{all scams} & \textsc{valid.} & \textsc{fn} & \textsc{rec.} \\ 
  \hline
IP contradicts location & 3030 (87\%) & 620 (87\%) &  44 (85\%) & 0.93 \\ 
  Suspicious language use & 2499 (72\%) & 507 (71\%) &  34 (65\%) & 0.93 \\ 
  IP address is a proxy & 2156 (62\%) & 433 (60\%) &  25 (48\%) & 0.94 \\ 
  Known scammer picture & 1379 (40\%) & 299 (42\%) &  17 (33\%) & 0.94 \\ 
  Known scammer details & 1368 (39\%) & 284 (40\%) &  13 (25\%) & 0.95 \\ 
  Self-contradictory profile & 1145 (33\%) & 242 (34\%) &  12 (23\%) & 0.95 \\ 
  IP location is suspicious & 968 (28\%) & 211 (29\%) &  22 (42\%) & 0.90 \\ 
  Mass-mailing other users & 761 (22\%) & 168 (23\%) &  10 (19\%) & 0.94 \\ 
  Picture contradicts profile & 261 (7\%) &  55 (8\%) &   4 (8\%) & 0.93 \\ 
   \hline
\end{tabular}
\label{tab:justifications}
}
\end{table}

Certain observations can be made. Firstly, on overall justification proportions
across scam profiles, we can see that examination of the geolocation of a
scammer's IP address is a heavily relied-on method for moderators, with
contradictions between this and the profile's stated location being a
justification listed for 87\% of all scam profiles. The next most common
justification was that a profile uses suspicious language in its
self-description: expressions of this ranged from identification of ``Nigerian
wording'' to moderators recognizing text being reused from previous scams . 

Comparison of proportions between the overall dataset and validation set show
little deviation in justification proportion, demonstrating a lack of bias. By
comparing proportions within the false-negative profiles to those in the overall 
validation set, we may discern any systemic differences in identification rate. 

Most justifications show similar or lower proportions in the false-negative
profiles, indicating that the ensemble is either no worse than average within
these subcategories, or may be better than average. One category of
justifications alone showed worse performance for the ensemble---where the
human moderators judged that the IP-determined origin of the scammer was in a
country they deem suspicious (e.g., a West African nation). The recall of
profiles justified with this reason was 0.9, lower than average. IP address
information is not available for non-scam users in our dataset, so this
discrepancy cannot be fully investigated, but it might suggest that the
partially location-based demographics classifier is not yet matching expert
understanding of scam-correlated locations.

\subsection{Evasion}

From the previous section, we see that moderators heavily rely on certain
features, such as the IP address, that can easily be obscured.  Moderators also
check if the IP address is known to belong to a proxy.  Although this could
certainly be used as a feature in our system, scammers could use unregistered
proxies (as yet inexpensive) or compromised hosts in residential IP address
space to evade detection. 
 
In contrast, our system relies on a wide range of features that are more
difficult to evade, such as textual features\footnote{Prior work in
computational linguistics has shown that the combination of (un)consciously made
linguistic decisions is unique for each individual --- like a fingerprint or a
DNA profile \cite{van2005new, rashid2013analyzing}}; or features for which their
obfuscation might render a profile unattractive, such as the demographics or the
images. 
\claudia{we could compare our results with the low performance of authorship attribution models when they
are confronted with adversarial passages (incl. obfuscation and/or imitation), but, again, I am not sure
this is relevant. What do you think?}

A natural scammer response to this kind of profile detection could be to cease
hand-crafting unusually attractive or targeted profiles, and instead turn to cloning
the existing profiles of real users from different dating sites. By
preferentially cloning attractive profiles, they would retain the high
match-rate which enables contact with potential victims, and by using real users'
profile information they would avoid detection systems geared towards 
their own idiosyncratic profile elements. Scammers are already
partially engaging in this sort of behaviour when they re-use images of real
people taken from the web.

The solution to such a development will rely on the deployment of
profile-cloning detection systems, such as that described by Kontaxis et
al.~\cite{kontaxis2011detecting}, perhaps augmented by behavioural classifiers 
operating on e.g., the language used in messages.
\claudia{Should we mention here that our system can be retrained to include and detect such clones?}

\subsection{Comparison of Dating Site Elements}
\label{appendix:dating_site_comparability}

\begin{table}[t]
\setlength{\tabcolsep}{0.1cm}
\centering
\caption{Comparison of the profile elements used in our classification experiments
with availability of these elements on popular dating sites. (\checkmark : present; {\bf \%}: requires inference)}
\scalebox{0.7}{
\begin{tabular}{lcccccccc}
  \hline
\textsc{Site} & age & gender & ethn. & marital & occ. & location & image & descr. \\ 
  \hline
\emph{\url{datingnmore.com}} & \checkmark & \checkmark &\checkmark &\checkmark &\checkmark &\checkmark &\checkmark &\checkmark  \\ 
\url{match.com} & \checkmark & \checkmark & \checkmark & \checkmark & \checkmark & \checkmark & \checkmark & \checkmark \\
\url{okcupid.com} & \checkmark & \checkmark & \checkmark & \checkmark & {\bf \%} & \checkmark & \checkmark & \checkmark \\
\url{pof.com} & \checkmark & \checkmark & \checkmark & \checkmark & \checkmark & \checkmark & \checkmark & \checkmark \\
\url{eharmony.com} & \checkmark & \checkmark & \checkmark & \checkmark & \checkmark & \checkmark & \checkmark & \checkmark \\
\url{tinder.com} & \checkmark & \checkmark & & & & {\bf \%} & \checkmark & \checkmark \\
   \hline
\end{tabular}
\label{tab:dating_site_comparison}
}
\end{table}

The \url{datingnmore.com} site used as the source of data for our experiments is comparatively small,
and has a niche appeal due to its intensive moderation by experts in identifying online dating fraud. 
It is therefore worthwhile considering its comparability to other dating sites, as a first step
to understand the generalisability of our results.

Table~\ref{tab:dating_site_comparison} compares the features from the \url{datingnmore.com} profiles 
which were used in our classifier with the profile elements available on five market-leading dating sites.
Coverage was good. All three of the ensemble components would be able to operate across these sites. The
image and description profile elements are always supported, and at least some demographic information is 
always available. The dating site with the fewest profile elements in common with our features is \url{tinder.com}, 
which has a distinctive locality-based use case which may hinder the online dating fraud our system aims to detect.

\subsection{Limitations \& Deployment Considerations}


There are limitations to our work which must be borne in mind. Firstly, whilst
we have taken pains to make use of profile features which should be visible on
other dating platforms, we have not yet tested our classification approach on
profiles from other dating sites. It may be the case that scammers and/or real
users show different characteristics in different dating platforms, which would
limit the applicability \claudia{robustness? cf. cross-platform categorisation studies} 
of our method. We are currently
seeking other sources of user profile data to investigate this possibility.
More information on scammer/user traits could generally
inform ongoing research into---and prevention of---online dating fraud.

Secondly, our results show a number of false negative
classifications. Further inspection of the data on scammers suggests that
augmenting our approach with other classifiers---such as ones using geolocated IP
addresses or observations of on-platform behavior (e.g., messaging)---could
help capture these scammers where the public profile information is
inconclusive. 


Online dating sites could deploy a detection system such as ours on their premises at the profile
registration stage. Security administrators would then be responsible for validating and acting upon 
the output of the classification system. How dating sites can responsibly anticipate and respond to errors in
automated classification systems such as this is a point of policy on which
practical industry insight would be highly valuable. More generally, this raises the issue of accountability
under the EU General Data Protection Regulation, and the ``right to explanation'' of algorithmic decisions that 
significantly affect a user. 



In the case where our system is deployed locally, the implications of our 
decisions can have a paramount effect on the user. Suppose that our system 
predicts that a given scam profile is safe. This might 
give the user a false sense of security, and encourage them into beginning a 
relationship with less caution than they would have applied otherwise. 
\claudia{These last two sentences sound a bit informal.}
Designing a tool which protects users while minimizing the risk of blind trust 
is a challenging interface design problem, but one which is outside the scope
of this paper. 
\claudia{If such a tool is available to users, what stops a scammer from using it
to test it on his/her own profile first to see the output other users will get?}

\section{Related Work}
\label{sec:background}

Despite the rapidly increasing number of victims\footnote{According to research at the Chartered Trading Standard Institute in the
UK, the number of romance scam victims will more than triple by 2019 \cite{Brown2017}.}, previous work on  
online dating fraud is limited, focussing mainly on case studies~\cite{rege2009love}, interviews with online dating site
users about their security practices~\cite{obada2017break} and interview and
questionnaire-based psychological profiling of victims~\cite{buchanan2014online,whitty2013scammers,whitty2016online}.

Three recent studies provide insight into romance scammer strategies. The authors of~\cite{yen2016case} 
carried out a study on the personals section of 
Craigslist to identify common methods of
romance scammers responding to honeypot adverts. Alongside identifying
approaches outside of traditional trust-building relationship-oriented scammers
(including driving users to other platforms and delivering hooks for
premium-rate numbers), they observed scammers were mostly West African in
origin, and used scraped images of attractive women~\cite{yen2016case}.  Huang
et al.~\cite{huang2015quit} performed a large-scale study of dating profiles
that are used by scammers, covering 500,000 scam accounts from an anonymous
Chinese online dating site. They found that different types of scammers target
different audiences and that advanced scammers are more successful in
attracting potential victims' attention.  Finally, the authors
of~\cite{edwards2018conPro} describe geographic variation in dating fraud
profiles, and propose a set of methods to improve geolocation when attackers
are hiding behind proxies. While assisting in attribution of origins, these
methods cannot be used for scammer detection.

Although a number of solutions based on machine learning techniques already exist to detect 
malicious activity on online services (e.g., detection of spam
\cite{lee2010uncovering,stringhini2010detecting,wang2010don}, 
or false identities~\cite{rashid2013analyzing,magdy2017fake}), 
to our knowledge, no prior work has attempted to automatically detect romance scammers. One of the main reasons for this is that
the dynamics of dating websites make scam detection more difficult than in other domains, such as
email or social networking. The intended operation of a dating site is that previously unconnected users will
reach out and initiate contact with people they do not know, and so spontaneous,
unsolicited communications cannot be viewed as a reliable signal of malicious behavior. Activities that in
other areas might be considered suspicious---contacting many users, providing
false profile attributes, migrating conversations to other media---could all also be considered normal behavior amongst
dating site users~\cite{huang2015quit}. Moreover, romance scams are for the most part carried out by
humans, adapting to changing circumstances, and so approaches which rely on 
detecting bot-like behavior are similarly stymied.  
In this paper, we address these issues by analyzing the launching point for the scam---the user profile.


\section{Conclusions}
\label{sec:conclusions}

In this paper, we have presented the first framework systematizing the 
identification of false online dating personas. 
Our exploratory analysis identified the sugarcoated 
lures used on fraudulent profiles. By analyzing the prevalence of these 
traits with respect to legitimate profiles, we 
engineered a diverse discriminatory feature set, using state-of-the-art text and image 
processing from multiple profile segments. This feature set allowed us to develop a set of 
independent classification systems which adjust to the omission of profile details.

Our experimental results show that our system can accurately detect online
dating fraud profiles, with high precision. A case by case analysis of our results, however, indicates that there 
are certain false profiles that look genuinely real. For these cases, we have noted that other 
sources of information, such as the messages exchanged, could be very
informative. As future directions, we aim to more broadly examine the available data on
online dating fraud, seeking information actionable for enforcement and other countermeasures.
We also hope to explore the question of how, at a local level, interventions designed
to warn and protect users from scammers can avoid forming dependences that reduce awareness.

\appendix

\section*{Acknowledgements}

This work is supported by award \texttt{EP/N028112/1}
\emph{``DAPM: Detecting and Preventing Mass-Marketing Fraud (MMF)''}, from the UK
Engineering and Physical Sciences Research Council. 
This research would not be possible without the work of the operators and
moderators of the \url{scamdigger.com} and \url{datingnmore.com} websites, and
their commitment to transparently combatting online dating fraud.


\section*{Availability}

To enable replication and foster research in this area we release 
the code used to obtain our data, 
the processing steps taken to prepare it, 
and the implementations of each classifier
together with additional details of our results, all available online at
\PublicRepo.





\bibliographystyle{abbrv}
\bibliography{romancedigger}

\end{document}